\documentclass[%
 reprint,
superscriptaddress,
 amsmath,amssymb,
aps,
]{revtex4-1}

\usepackage{graphicx}
\usepackage{dcolumn}
\usepackage{bm}
\usepackage{xspace}
\newcommand{\bra}[1]{{\langle #1 |}}
\newcommand{\ket}[1]{{| #1 \rangle}}
\newcommand{\braket}[2]{{\langle #1 | #2 \rangle}}
\newcommand{\CT}{Cr$_2$Te$_3$\xspace}

\newcommand{\CrI}{Cr$_1$\xspace}
\newcommand{\CrII}{Cr$_2$\xspace}
\newcommand{\CrIII}{Cr$_3$\xspace}

\newcommand{\kv}{{\bf k}}

\newcommand{\Rv}{{\bf R}}
\newcommand{\ep}{{\epsilon}}

\begin{document}

\preprint{APS/123-QED}

\title{Evolution of Berry Phase and Half-Metallicity in Cr$_2$Te$_3$ in Response to Strain, Filling, 
Thickness, and Surface Termination}

\author{Sohee Kwon}
\email{skwon054@ucr.edu}
\affiliation{Laboratory for Terahertz $\&$ Terascale Electronics (LATTE), Department of Electrical and Computer Engineering, University of California-Riverside, Riverside, CA, 92521, USA}
\author{Yuhang Liu}
\affiliation{Laboratory for Terahertz $\&$ Terascale Electronics (LATTE), Department of Electrical and Computer Engineering, University of California-Riverside, Riverside, CA, 92521, USA}
\author{Hang Chi}
\affiliation{Department of Physics, University of Ottawa, Ottawa, ON, K1N 6N5, Canada}
\affiliation{Nexus for Quantum Technologies, University of Ottawa, Ottawa, ON, K1N 6N5, Canada}
\author{Gen Yin}
\affiliation{Department of Physics, Georgetown University, Washington, DC 20057, USA}
\author{Mahesh R. Neupane}
\email{mahesh.r.neupane.civ@army.mil}
\affiliation{Laboratory for Terahertz $\&$ Terascale Electronics (LATTE), Department of Electrical and Computer Engineering, University of California-Riverside, Riverside, CA, 92521, USA}
\affiliation{DEVCOM Army Research Laboratory, 2800 Powder Mill Rd, Adelphi, MD, 20783, USA}
\author{Roger K. Lake}
\email{Corresponding author: rlake@ece.ucr.edu}
\affiliation{Laboratory for Terahertz $\&$ Terascale Electronics (LATTE), Department of Electrical and Computer Engineering, University of California-Riverside, Riverside, CA, 92521, USA}

%
%

\date{\today}

\begin{abstract}
Cr$_2$Te$_3$ is a ferromagnetic, quasi-two-dimensional layered material with perpendicular magnetic anisotropy, 
strong spin-orbit coupling, and non-trivial band topology.
The non-trivial topology results in an intrinsic anomalous Hall conductivity (AHC) that switches sign under
filling and biaxial strain.
Thin films can exhibit half metallicity.
Using density functional theory combined with maximally localized Wannier functions, we reveal the physical 
origins of the sensitivity of the sign of the AHC to strain and filling, and we determine the effect  
of surface termination on the half metallicity.
We find that thin films terminated on the Te layers are the most energetically stable, 
but only the thin films terminated on both sides with the partially occupied Cr layers
are half metals.
In bulk \CT,
the sensitivity of the sign of the AHC to strain and filling
results from the complex Fermi surface comprised of three bands.
Filling of local minima
and bands near anti-crossings
alters the local Berry curvature
consistent with the negative to positive switching of the AHC.
Similarly, strain depopulates a local minimum, shifts a
degenerate point closer to the Fermi energy, and causes
two spin-orbit split bands to reverse their order. 
These findings provide a physical understanding of the evolution
of the Berry phase, AHC, and half-metallicity in \CT.

\end{abstract}

\maketitle


\section{Introduction}
The magnetic structure of bulk \CT has been extensively investigated for decades
\cite{1970_Andresen_Acta_Chem_Scand,
1971_Cr2Te3_Hashimoto_JPSJ,
1975_Cr2Te3_Watanabe_SSCom,
1989_E-k_M_Cr2Te3_Dijikstra_JPCM,
1993_Mag_Res_Konno_JJAP,
2007_Correlation_Effects_Cr2Te3_JAP,
2012_Intrinsic_Ex_Bias_JAP}.
%
More recently, interest has shifted to thin films of \CT
\cite{2014_gating_PSSC,
2019_Burn_SciRep,
2019_MBE_Thin_Films_Tune_Tc_ACS_App_NMater,
2021_Cov_2D_Cr2Te3_FM_MatResLett,
2019_Cr2Te3_TI_SciRep,
2021_Cr2Te3_Bi2Te3_NLett,
2022_strain-sensitive_Zhong_NRes,
2023_Stonor_to_Local_Zhong_NComm}.
Due to the van der Waals coupled CrTe$_2$ layers and unique self-intercalated Cr layers, 
Cr$_{1+x}$Te$_2$ is a promising candidate for modulating both its magnetic properties 
and topological properties \cite{Fujisawa2022}.

Half-metallicity is an important feature in high performance spintronic devices.
In a half metal, there is a band gap in one spin channel and 
metallic behavior in the other. 
As a result, the electrons at the Fermi level are fully spin polarized.
Within the Cr-Te family, half-metallic ferromagnetism was reported in bulk zincblende 
CrTe \cite{half-metal_CrTe_2003}.
There has been little focus on the effect of surface termination on the stability and half-metallicity in few-layer films. 
Bian $et$ $al.$ \cite{2021_Cov_2D_Cr2Te3_FM_MatResLett} 
found that Cr$_{1+\delta}$Te$_2$ thin films terminated by Te layers have in-plane lattice contraction compared to that of the bulk and to that of Cr terminated films, and they concluded that few-layer films terminated with Te layers were unstable.
They found that Cr terminated \CT films 
with thicknesses ranging from
monolayer to 4 layers exhibit half-metallic behavior.
%

The anomalous Hall conductivity (AHC) of various Cr-Te compounds has been actively studied.
Huang $et$ $al.$ \cite{2021_CrTe2_colossalAHE_acsnano} reported an AHC of 67 k$\Omega$cm, 
and a corresponding anomalous Hall angle of 5.5\%
in a 170 nm thick 1T-CrTe$_2$ flake at $T$=3 K.
It was determined that
the large AHC originated from extrinsic skew scattering rather than from the intrinsic Berry phase.
Fujisawa $et$ $al.$ \cite{Fujisawa2022} 
investigated  Cr$_{1+\delta}$Te$_2$ thin films 
grown by molecular beam epitaxy (MBE) and
tuned by the stoichiometry ratio in the range 0.3 $< \delta <$ 0.8.
Angle-resolved photoemission spectroscopy measurements showed a relatively rigid
shift of the Fermi energy ($E_F$) as a function of stoichiometry of
$\Delta E_F$/$\Delta\delta \sim$ 1.54 eV/atom.
The sign of the AHC changed from positive to negative as $\delta$ varied from 0.33 to 0.47.
The experimental values of the AHC qualitatively matched the calculated intrinsic AHC values, including the change of sign, 
although the calculated values were two orders of magnitude greater than the experimental ones.

Chi $et$ $al.$ \cite{2023_our_NatComm} reported strain tunable anomalous Hall conductivity in \CT thin films.
The growth of various thicknesses of \CT films on Al$_2$O$_3$ and SrTiO$_3$
resulted in different levels of average strain. 
A sign reversal of the AHC was observed under 
compressive strain both experimentally and theoretically.
The quantitative value of the AHC, 
calculated from the intrinsic Berry phase of the unstrained crystal, agreed with the experimental value.
Furthermore, under the application of compressive biaxial strain, 
the calculated intrinsic AHC was found to change sign, consistent with the
experimental data.

Guillet $et$ $al.$ \cite{2023_arxiv_Cr2Te3_on_2D_exp_DFT} demonstrated MBE growth of 
5 monolayers of \CT on sapphire and on various 2D substrates: 
graphene, WSe$_2$, and Bi$_2$Te$_3$.
A sign change in the AHC was observed for samples on sapphire.
Theoretically,
a sign reversal of the AHC occurred as $E_F$ was shifted by 10 meV with
respect to its equilibrium value.
Possible origins of the AHC sign reversal were suggested to be strain dependence, 
thermal broadening upon heating, and charge transfer due to the substrate effect.
The authors claimed that the AHC was primarily from the intrinsic effect, however, the Berry phase analysis was not discussed in detail.
The effect of strain was also studied theoretically by Gebredingle $et$ $al.$  \cite{2022_Cr2X3_DFT_ApplNanoMat} for Cr$_2$X$_3$ (X = S, Se, Te),
although its affect on the AHC was not considered.
In this paper, we determine the effect of surface
termination on thin-film stability and half-metallicity.
We then show how strain and band filling affect the
electronic band structure and the Berry curvature, which
is the physical origin of the sign reversal of the intrinsic AHC in bulk \CT.
%

\section{Methodology}
%
%
\begin{figure}[htbp]
\includegraphics[width=0.5\textwidth]{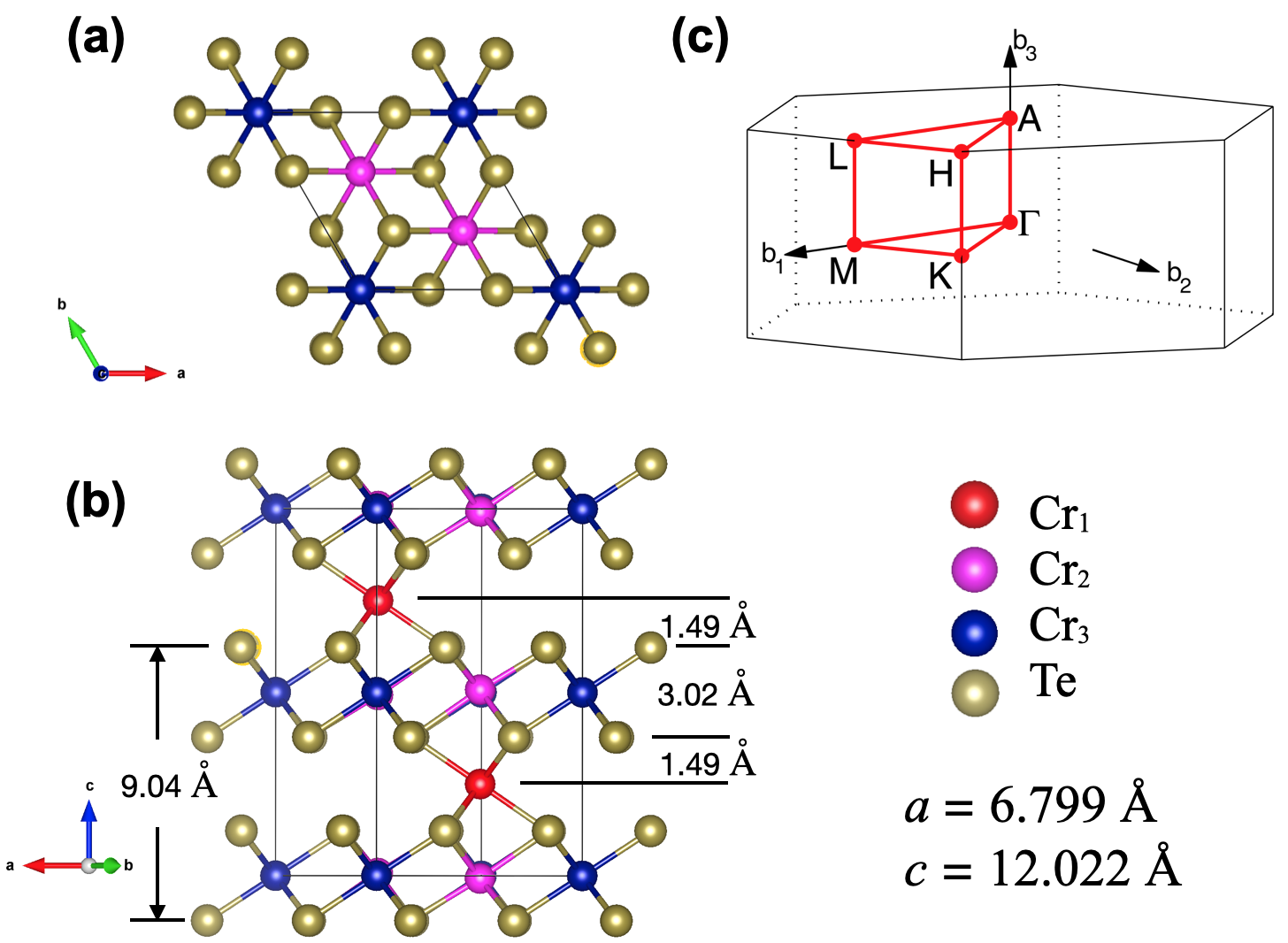}
\caption{(a) Top (001) and (b) side (210) views of bulk Cr$_2$Te$_3$. 
 Interlayer distances are shown. 
 The vertical and horizontal
 thin lines in (a,b) show the unit cell from the (001) and (210) directions.
 The legend provides the labels for the
 3 non-equivalent Cr atoms in the unit cell.
 The values for the $a$ and $c$ lattice constants and
 the inter-layer distances of the fully
 relaxed structure are also shown.
 (c) The Brillouin zone and the high symmetry lines used for plotting
band structure and Berry phase.}
\label{fig_structure}
\end{figure}
The first-principles electronic structure calculations are carried out within the projector augmented-wave method, 
as implemented in the Vienna $ab$ $initio$ simulation package (VASP) \cite{VASP}.
The generalized gradient approximation is used to describe the exchange-correlation functional as 
parametrized by Perdew, Burke, and Ernzerhof (PBE) \cite{PBE}.
The van der Waals interaction is included
using the DFT-D3 method of Grimme {\em et al.} with zero-damping \cite{grimme2010consistent}.
We apply biaxial strain and relax the atomic positions within the unit cell.
After relaxation, the forces acting on the ions are less than $5 \times 10^{-3}$ eV/\AA, 
and the change in the total energy between two ionic relaxation steps is smaller than $10^{-6}$~eV.
The high symmetry $k$-path has been chosen by the AFLOW software \cite{Aflow}.
With a cutoff energy of 520 eV, we use a converged Monkhorst Pack 8$\times$8$\times$4 $k$-grid for 
geometric relaxation and a $\Gamma$-centered 12$\times$12$\times$6 $k$-grid for the electronic calculation including spin-orbit coupling (SOC).

%
Figure \ref{fig_structure}(a,b) displays a top and cross-sectional view of the 
hexagonal unit cell of bulk Cr$_2$Te$_3$ consisting of 8 Cr atoms and
12 Te atoms.
The Brillouin zone, and the high symmetry paths are shown in Fig.  \ref{fig_structure}(c).
The space group of Bulk \CT is 
$P\bar{3}$1c (No.163)
with sixfold in-plane symmetry \cite{2023_our_NatComm}.
The unit cell consists of alternating layers in which all of the metal sites 
are either fully occupied with Cr atoms 
or partially occupied with Cr atoms.
The partially occupied layers are also referred to as `vacancy layers'
or `interstitial layers'.
There are three inequivalent Cr atoms in the unit cell.
The Cr atoms in the two vacancy layers are equivalent and labeled as \CrI.
The fully occupied layers contain two inequivalent Cr atoms.
The Cr atoms directly above and below the \CrI atoms are labeled as \CrII.
The Cr atoms at the corners and edges of the fully occupied layers are 
labeled as \CrIII.
In the unstrained state, the optimized lattice constants of bulk 
Cr$_2$Te$_3$ are $a= 6.799$ \AA\ and $c= 12.022$ \AA\ .
Structural stability is checked by performing phonon spectra and molecular dynamics
calculations.
The phonon dispersion is calculated using the second-order interatomic force constants (IFCs) 
and the finite displacement method as implemented in Phonopy \cite{phonopy}.
$Ab$ $initio$ molecular dynamics (AIMD) calculations, as implemented in VASP, 
are performed applying a canonical ensemble with a Nose-Hoover thermostat at room temperature.

%
\begin{figure}[htbp]
	\includegraphics[width=0.45\textwidth]{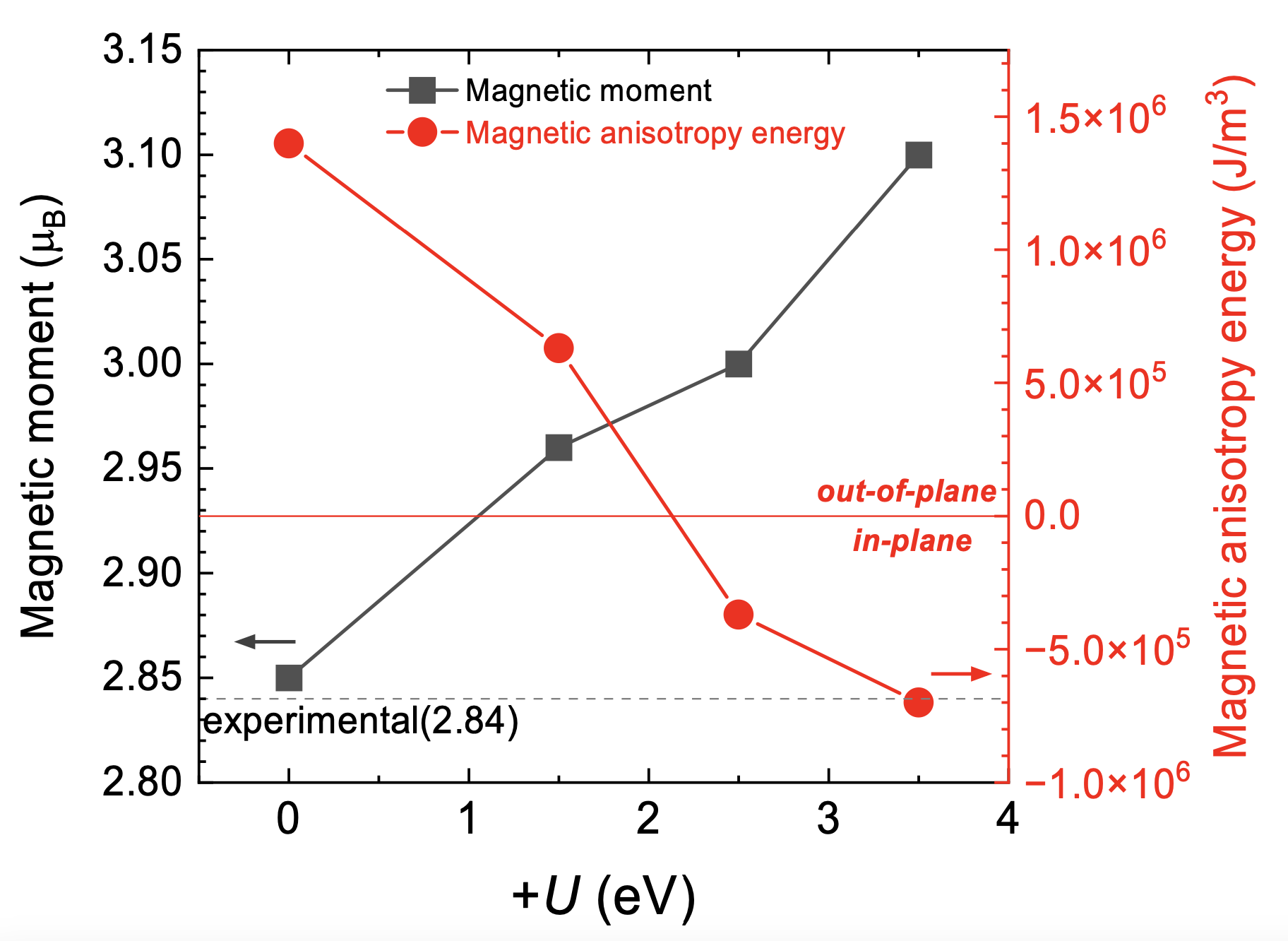}
	\caption{Average magnetic moment per Cr atom (left axis) and magnetic anisotropy (right axis) as a function of $+U$ parameters of the Cr atom.
Magnetic moments are in units of Bohr magnetons ($\mu_B$).}
	\label{fig_MAE_U}
\end{figure}
To choose the Hubbard $U$ parameter for the Cr atoms, 
we investigate values of $U$ ranging from 0 to 4 eV.
As the value of $U$ increases, 
the average magnetic moments per Cr ion increase monotonically from 
2.85 $\mu_B$ to 3.10 $\mu_B$ 
while the magnetic anisotropy energy (MAE) decreases monotonically  from
1.4$ \times 10^6$ J/m$^3$ to -7 $\times 10^5$ J/m$^3$.
%
%
%
The values are plotted in Fig. \ref{fig_MAE_U}.
The MAE
changes sign at $\sim U = 2$ eV 
at which point out-of-plane anisotropy (easy-axis along the $c$ axis) gives way to easy-plane anisotropy
in the $a$-$b$ plane.
%
%
According to experimental results \cite{2023_our_NatComm}, 
bulk Cr$_2$Te$_3$ has a strong out-of-plane magnetic anisotropy, and a magnetic moment 
per Cr atom of 2.82 $\mu_B$ \cite{1970_Andresen_Acta_Chem_Scand}.
%
With $U$= 0 eV, we obtain magnetic moments of 2.85 $\mu_B$ per Cr atom 
and an out-of-plane MAE of 0.5 meV per Cr atom.
We also calculated the AHC using a value of $U=2$ eV, and
the result differs from the experimentally measured value by
a factor of 10 as shown in Table \ref{tab:U} of Appendix
\ref{k-convergence}.
Therefore, $U= 0$ is used for all of the calculations in this paper.
%

The WANNIER90 package \cite{1997_MLWF_PRB,2001_MLWF_PRB,2014_Wannier90,2020_Wannier90} is used 
to transform the Hamiltonian into the Wannier basis 
and to calculate topological properties, 
such as the Berry curvature and the AHC, 
with a denser $k$-mesh of 100$\times$100$\times$100.
The AHC $k$-convergence test results are shown in Appendix \ref{k-convergence}.
We employ maximally localized Wannier functions (MLWFs) including Cr $d$-orbitals 
and Te $p$-orbitals.
For MLWF calculations, we set the magnetization along the $z$-axis and include SOC.
With 5 $d$ orbitals $\times 2$ spins per Cr atom
and 3 $p$ orbitals $\times 2$ spins per Te atom, 
the resulting Wannier basis consists of 152 orbitals per unit cell.
%
The Hamiltonian matrix elements $\bra{m,0}H\ket{n,\Rv}$ and the Wigner-Seitz
cell degeneracies $N_D(\Rv)$ determined from Wannier90 are used to construct
the Hamiltonian in the Bloch-sum basis, 
$H^W_{mn}(\kv)=\sum_{\Rv} \bra{m,0}H\ket{n,\Rv} e^{i\kv \cdot \Rv} / N_D(\Rv)$.
The eigenvalues of $H^W_{mn}(\kv)$ are used
to generate the multiple 
slices of the Fermi surface with each slice calculated on a 400$\times$400
$k$-grid.

For current in the $a$-$b$ plane, the intrinsic AHC is governed by the  
$z$ component of Berry curvature ($\Omega^z(k)$), which can be expressed as \cite{Yao2004},
\begin{align}
\Omega^z_n(\kv) & =	 \sum_{n'\neq n}	
\left[
\frac{\langle\psi_{n\kv}|v_x|\psi_{n'\kv}\rangle\langle\psi_{n'\kv}|v_y|\psi_{n\kv}\rangle}{(\omega_{n'\kv}-\omega_{n\kv})^2} 
\right.
\nonumber \\
& \;\;\;\;\;\;\;\;\;\;\;\;\;\;\;\;
- \left. \frac{\langle\psi_{n\kv}|v_y|\psi_{n'\kv}\rangle\langle\psi_{n'\kv}|v_x|\psi_{n\kv}\rangle}{(\omega_{n'\kv}-\omega_{n\kv})^2} 
\right]
\label{eq:Omega-n-z-1}
\\
& = -2 {\rm Im}
\sum_{n'\neq n}
\frac{\langle\psi_{n\kv}|v_x|\psi_{n'\kv}\rangle\langle\psi_{n'\kv}|v_y|\psi_{n\kv}\rangle}{(\omega_{n'\kv}-\omega_{n\kv})^2} ,
\label{eq:Omega-n-z-2}
\end{align}
where $v$ is the velocity operator, $n$ is the band index, and the band energy at 
wavevector $\kv$ is $E_{n\kv}=\hbar\omega_{n\kv}$.
The intrinsic AHC is determined from the sum of $\Omega^z_n(\kv)$ over the occupied bands,
\begin{equation}
\Omega^z(\kv) = - \sum_n f_n(\kv) \Omega^z_n(\kv),
\label{eq:Omega-z}
\end{equation}
where $f_n(\kv)$ is the occupation of band $n$ at wavevector $\kv$.
The intrinsic AHC is obtained by integrating $\Omega^z(\kv)$ over the Brillouin zone.
\begin{equation}
\sigma_{xy} = - \frac{e^2}{\hbar} \int_{BZ} \frac{d^3k}{(2\pi)^3} \Omega^z(\kv).
\end{equation}

\section{Results and Discussion}
%
%
\subsection{Structural Stability and Half-metallicity in Thin Films Cr$_{1+\delta}$Te$_2$}

\begin{figure*}[htbp]
\includegraphics[width=0.8\textwidth]{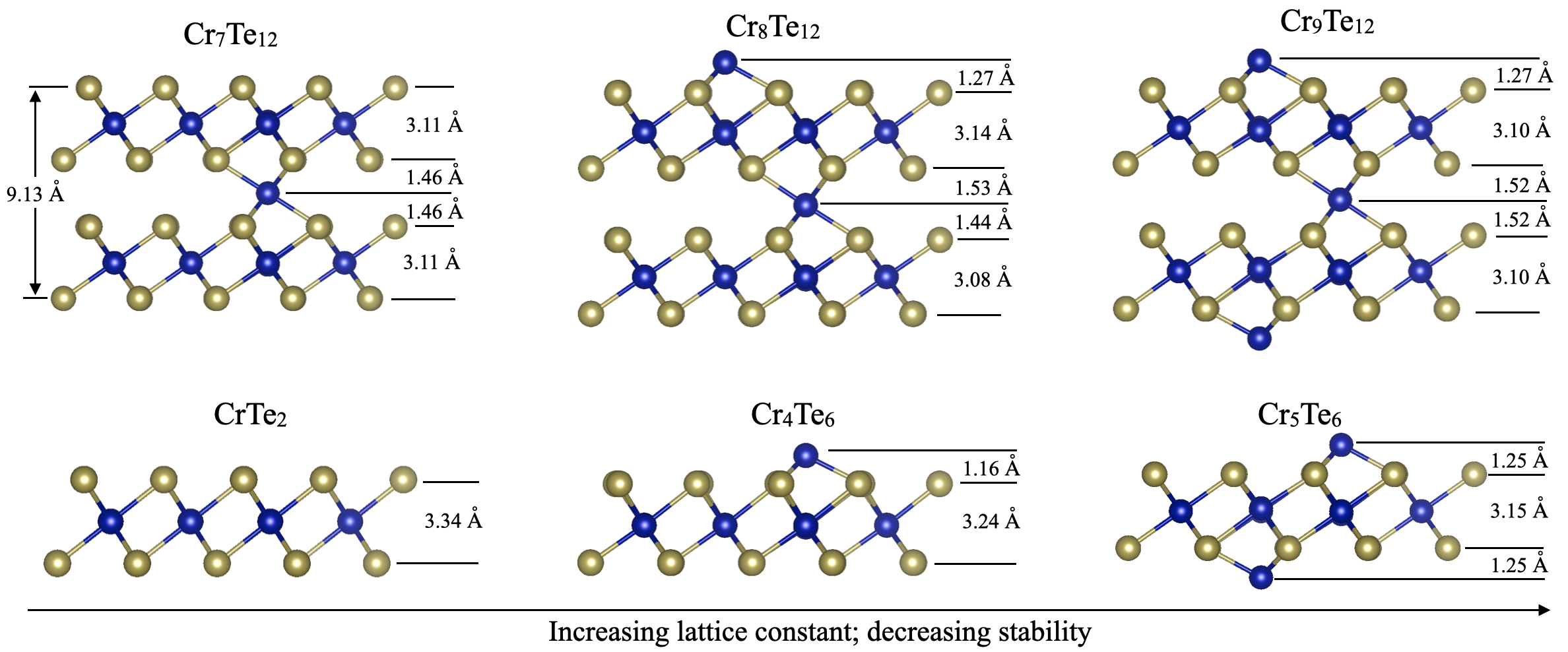}
\caption{
Side views of the thin-film structures arranged from
left to right in order of increasing in-plane lattice
constant and decreasing stability. 
Inter-layer distances are shown.
}
\label{fig_Eform}
\end{figure*}

The formation energies, lattice constants, 
and magnetic properties of finite slab structures are affected by the surface termination. 
We first consider the effect of surface termination on the structural stability of few-layer films. 
We calculate the formation energies and in-plane 
lattice constants for different surface terminations and layer numbers,
and we also calculate the phonon dispersions and perform $ab$ $initio$ molecular
dynamics to confirm structural stability at room temperature.
The results from the phonon and molecular dynamics calculations are displayed in 
Appendix \ref{app:phonon_disp_MD}.

\begin{table*}[htbp]
\renewcommand\arraystretch{1.25}
\caption{
In-plane lattice constant $a$, 
vertical height $h$ with the equivalent bulk distance in parentheses, 
formation energy $E_{\rm form}$, 
and formation energy per atom for the slab structures in 
Fig. \ref{fig_Eform} and for bulk Cr$_{1+\delta}$Te$_2$.
The quantities $\ep_{\parallel}$, 
$\ep_{\perp}$,
and
$\Delta \Omega / \Omega_{\rm bulk}$,
are the 
in-plane,
out-of-plane,
and
volumetric percent changes with respect to the bulk values.
}
\begin{ruledtabular}
\begin{tabular}{l| c c c c c c c}
System     & 1L CrTe$_2$     & 1L Cr$_4$Te$_6$     & 1L Cr$_5$Te$_6$     & 2L Cr$_7$Te$_{12}$     & 2L Cr$_8$Te$_{12}$     & 2L Cr$_9$Te$_{12}$   & bulk Cr$_2$Te$_3$ \\ \hline
Cr(Te) per u.c. & 3(6) & 4(6) & 5(6) & 7(12) & 8(12) & 9(12) & 8(12) \\ 
In-plane $a$ (\AA) & 6.350 & 6.638 & 6.849 & 6.668 & 6.726 & 6.812 & 6.799 \\
$\ep_{\parallel}=\frac{a-a_{\rm bulk}}{a_{\rm bulk}}$(\%) & -6.6 & -2.4 & 0.7 & -1.9 & -1.1 & 0.2 & 0\\ 
Vertical $h$ (\AA) & 3.34(3.02) & 4.40(4.52) & 5.66(6.01) & 9.13(9.04) & 10.47(10.53) & 11.74(12.02) & 12.02\\
$\ep_{\perp}=\frac{h-h_{\rm bulk}}{h_{\rm bulk}}$(\%) & 10.4 & -2.7 & -5.8 & 1.0 & -0.6 & -2.3 & 0\\
$\Delta \Omega/\Omega_{\rm bulk}$(\%) & -3.8 & -7.2 & -4.4 & -2.4 & -2.7 & -2.0 & 0\\
$E_{\rm form}$ (eV) & -3.270 & -2.240 & -1.098 & -8.612 & -7.427 & -6.335 &  -10.818 \\
$E_{\rm form}$/atom (meV) & -363 & -224 & -100 & -453 & -371 & -302 &  -541 \\
\end{tabular}
\end{ruledtabular}
 \label{table_Eform}
\end{table*}

We investigated three different types of terminations of the
top/bottom Cr$_{1+\delta}$Te$_2$ slab surfaces: Te/Te, Cr/Te, and Cr/Cr.
Figure \ref{fig_Eform} shows the structures considered 
for monolayer and bilayer Cr$_{1+\delta}$Te$_2$ 
with different surface terminations.
The inter-layer distances are also shown.
The structures are arranged from left to right in the order of 
decreasing stability.
Table \ref{table_Eform} shows 
in-plane lattice constants, vertical heights,
the in-plane, perpendicular, and hydrostatic strains
(defined with respect to the bulk distances),
formation energies, and formation energies per atom.

The formation energy is calculated from
\begin{equation}
E_{\rm form}=E_{\rm total}-(nE_{\rm Cr}+mE_{\rm Te})
\label{eq:Eform}
\end{equation}
where $n$($m$) is number of Cr(Te) atoms in a unit cell and
$E_{\rm Cr}$($E_{\rm Te}$) is the single atom energy obtained
from the total energy per atom of the bulk crystal of Cr(Te).
The calculated values are $E_{\rm Cr} = -9.528$ eV and $E_{\rm Te} = -3.144$ eV.
The formation energy per atom is $E_{\rm form}/(n+m)$.
The formation energies are defined such that more negative values correspond
to greater stability.
The results in Table \ref{table_Eform} show that the structures terminated on the Te
layers have the lowest formation energies in both monolayer and bilayer structures.
Thus, the order of stability of both monolayer and bilayer films, 
arranged in decreasing order, 
is Te/Te-terminated $>$ Te/Cr-terminated $>$ Cr/Cr-terminated.
While the Te/Te terminated films are most stable, 
all terminations are stable at room temperature, as shown by the molecular
dynamics calculations in Appendix \ref{app:phonon_disp_MD}. 
Thus, it is likely that all terminations can be found in experiments. 

The changes in the optimized in-plane lattice constants and vertical
inter-layer distances also have monotonic dependencies on the layer
terminations, but in opposite directions.
The in-plane lattice constants of the Te/Te terminated films 
(CrTe$_2$ and Cr$_7$Te$_{12}$) decrease,
and the vertical heights between the highest and lowest Te layers increase with
respect to the bulk values.
For the Cr/Cr terminated films (Cr$_5$Te$_6$ and Cr$_9$Te$_{12}$),
the opposite is true.
The in-plane lattice constants slightly increase, and the vertical heights
significantly decrease with respect to that of the bulk.
In both of the above cases, the change in the in-plane and out-of-plain 
dimensions have opposite signs, as would be expected to preserve volume.
For the Janus type Te/Cr termination, the changes in both
the in-plane and out-of-plane dimensions
are negative. 
This could be seen as the behavior of an auxetic material with a negative
Poisson's ratio 
\cite{2017_Neg_Pois_1T_TMDCs_Ncomm,
2020_2D_Neg_Poisson_JMatChemC,
2023_Neg_Pois_Janus_APL}, 
however, as we discuss further below, this behavior is better
viewed as a surface effect due to the Cr termination.
%

These trends are quantified in terms of the 
in-plane, out-of-plane, and volumetric percent changes
with respect to the bulk values:
$\ep_{\parallel}$, $\ep_\perp$, and $\Delta \Omega/\Omega_{\rm bulk}$. 
The definitions are given in the left column of Table \ref{table_Eform}.
While we are using the symbols generally used for strain,
we emphasize that these quantities are simply percent changes 
with respect to the equivalent bulk quantities.
They are not actual strain, since each slab is fully relaxed.
The in-plane percent change $\ep_{\parallel}$ is large and negative 
for the Te/Te terminated films, 
and it becomes small and positive for the Cr/Cr terminated films.
The out-of-plane percent change $\ep_\perp$ is large and positive for the 
Te/Te terminated films,
and it becomes large and negative for the Cr/Cr terminated films.
For all structures, the volumetric percent change
$\Delta \Omega / \Omega_{\rm bulk} \approx 2 \ep_{\parallel} + \ep_{\perp}$,  
is negative.
For the mono-layer (1L) structures, the magnitude of the volumetric percent change
is smallest for Te/Te termination, 
because of the compensating signs and magnitudes of $\ep_{\parallel}$
and $\ep_\perp$.
For the bilayer (2L) structures, the magnitude of the volumetric percent changes
for Cr/Cr and Te/Te termination are comparable, -2.0 and -2.4, respectively.  
For both 1L and 2L structures, the magnitudes of volumetric percent changes
are maximum for the Te/Cr terminated structures, since the signs of 
$\ep_\parallel$ and $\ep_\perp$ are both negative.

There are two competing trends that determine the percent changes
in the vertical heights of the slabs with respect to that of
the bulk: the vertical expansion of the CrTe$_2$ layers and the 
large contraction of the vertical distances of the surface Cr layers.
The thickness of the bulk CrTe$_2$ layer is 3.02 {\AA}
(see Fig. \ref{fig_structure}). 
For the monolayers 
(bottom row of Fig. \ref{fig_Eform}),
the vertical expansion is largest for the single 
CrTe$_2$ monolayer (3.34 \AA), and it decreases
as Cr atoms are added to the top and bottom surfaces.
For the bilayers, all of the CrTe$_2$ layers are slightly expanded
compared to that of the bulk, but the trend with respect to the
addition of Cr surface layers is weak or nonexistent.
The qualitative behavior of the expansion of the CrTe$_2$ layer is identical to that observed when
comparing bulk 1T-CrTe$_2$ to bilayer and monolayer 1T-CrTe$_2$ \cite{Yuhang_CrTe2,Yuhang_CrTe2_data}. 
\begin{figure*}[htbp]
\includegraphics[width=1\textwidth]{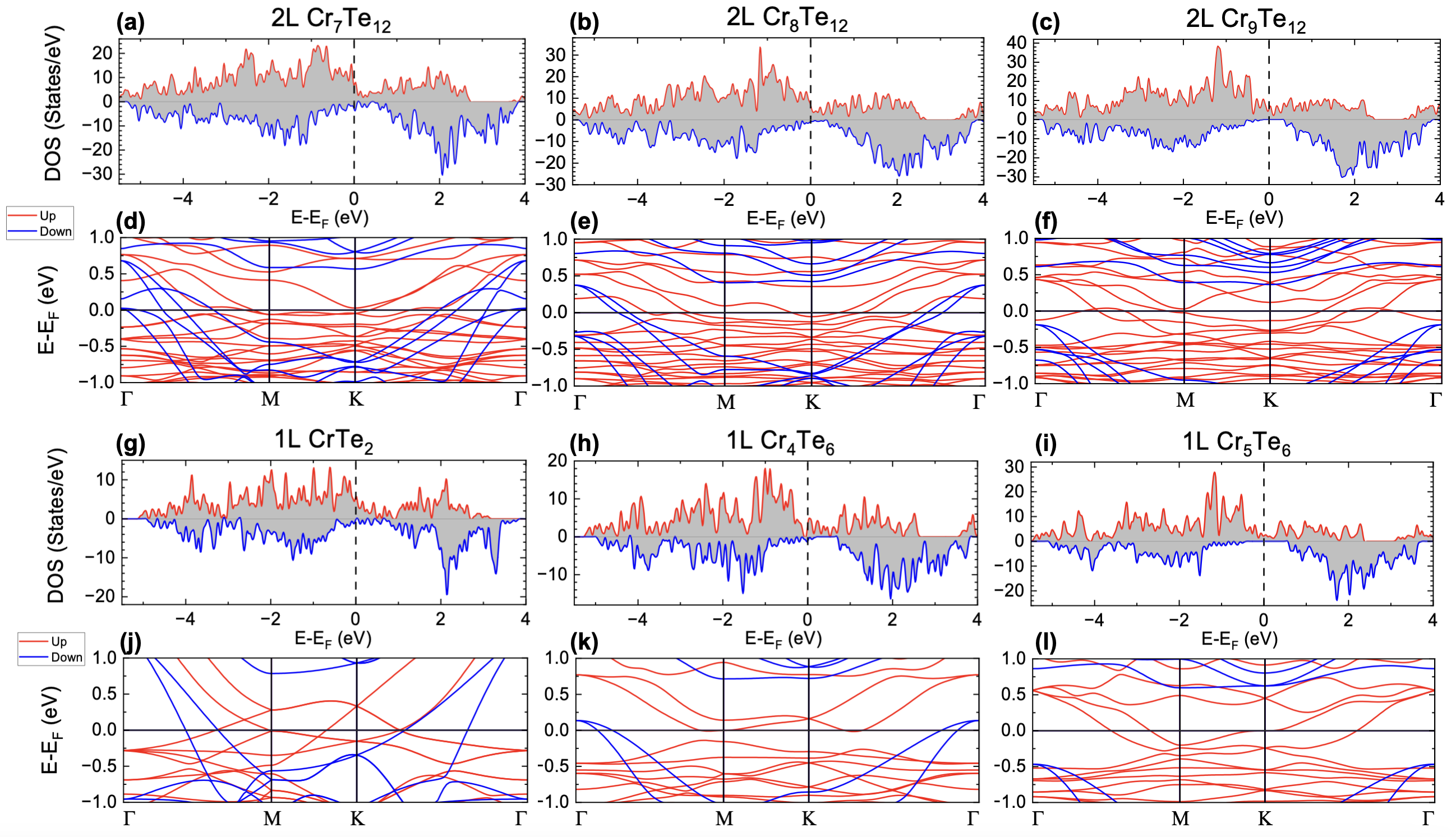}
\caption{
 Spin-resolved density of states and electronic band structures of differently terminated Cr$_{1+\delta}$Te$_2$ few-layer films. 
 Red (blue) indicates the majority (minority) spin channel for both the 
 density of states and the electronic band structure ($E$-$k$) plots.} 
 The density of states of Te/Te, Cr/Te, and Cr/Cr terminated bilayers are shown in (a-c), respectively,
 and the corresponding $E$-$k$ plots are shown in (d-f). 
 The spin-resolved density of states of Te/Te, Cr/Te, and Cr/Cr terminated monolayers are shown in 
 (g-i), and the corresponding $E$-$k$ plots are shown in (j-l).
\label{fig_DOS}
\end{figure*}

The opposing trend is the reduction of the distance
of the partially occupied Cr surface layer from the Te layer.
In the bulk, the Cr$_1$-Te inter-layer distance is 1.49 {\AA}
(see Fig. \ref{fig_structure}).
In the Cr$_8$Te$_{12}$ and Cr$_9$Te$_{12}$ bilayers, 
that distance shrinks by -15\% to 1.27 \AA. 
In the monolayers, the shrinkage is slightly larger.
It is the large vertical contraction of this surface Cr
layer distance that causes $\ep_\perp$ to be negative for all
of the structures terminated on at least one surface with Cr, i.e.
the Cr/Te and Cr/Cr structures.
This is the reason for the unusual trend present in the 1L and 2L
Cr/Te terminated structures in which both $\ep_\parallel$ and $\ep_\perp$
are negative. 
Another distinguishing feature of Cr$_{1+\delta}$Te$_2$ thin films is half-metallicity.
The density of states of differently terminated thin films are depicted in Fig. \ref{fig_DOS}.
The Te/Te-terminated monolayer CrTe$_2$ (Fig. \ref{fig_DOS}(g)) and bilayer Cr$_7$Te$_{12}$ (Fig. \ref{fig_DOS}(a)) have both majority and minority-spin states at the Fermi-level.
In the Cr/Te-terminated monolayer and bilayer Cr$_2$Te$_3$ structures, 
the density of minority-spin states at the Fermi-energy is low, 
as shown in Fig. \ref{fig_DOS}(b,h).
The electronic band structures of these monolayer and bilayer 
Cr/Te terminated films in Fig. \ref{fig_DOS}(e,k) show that two minority-spin hole 
bands cross the Fermi-level near $\Gamma$.
Adding another partially occupied Cr layer on the bottom surface, 
such that both surfaces are Cr-terminated, 
the thin films become half-metals as shown in Fig. \ref{fig_DOS}(c,i).
The minority-spin hole bands at $\Gamma$ are now 0.19 eV (0.47 eV) below the 
Fermi energy for the 2L (1L) structures, 
and the bands crossing the Fermi level are all majority carrier bands. 
The spin gap of monolayer Cr$_5$Te$_{6}$ is 1.07 eV and that of bilayer 
Cr$_9$Te$_{12}$ is 0.55 eV, as shown in Fig. \ref{fig_DOS}(f,l).
The Cr added to the Te terminated surfaces acts as a donor-dopant, since it raises the Fermi 
level with respect to the existing bands, and it also increases the minority spin gap
between the minority spin valence band at $\Gamma$ and the minority spin conduction band
at M. 
These results are consistent with the electronic structure of bulk CrTe, which is a half-metallic ferromagnet \cite{half-metal_CrTe_2003}, since adding extra Cr layers on the surfaces of the \CT thin films makes their stoichiometry closer to that of CrTe.

\subsection{Anomalous Hall Conductivity and Berry Curvature in Bulk \CT}
\label{sec:Berry_Curv_AHC}
The sign of the AHC is sensitive to both band filling and strain.
We will show that the origin of the sensitivity 
results from the filling, emptying, and shifting of 
three bands that cross the Fermi level.
We now calculate 
the changes in the AHC of bulk \CT
resulting from filling and strain, and then 
analyze the changes by determining the effect of filling and strain on 
the electronic band structure and Berry curvature.
%
\begin{figure}[htbp]
\includegraphics[width=0.37\textwidth]{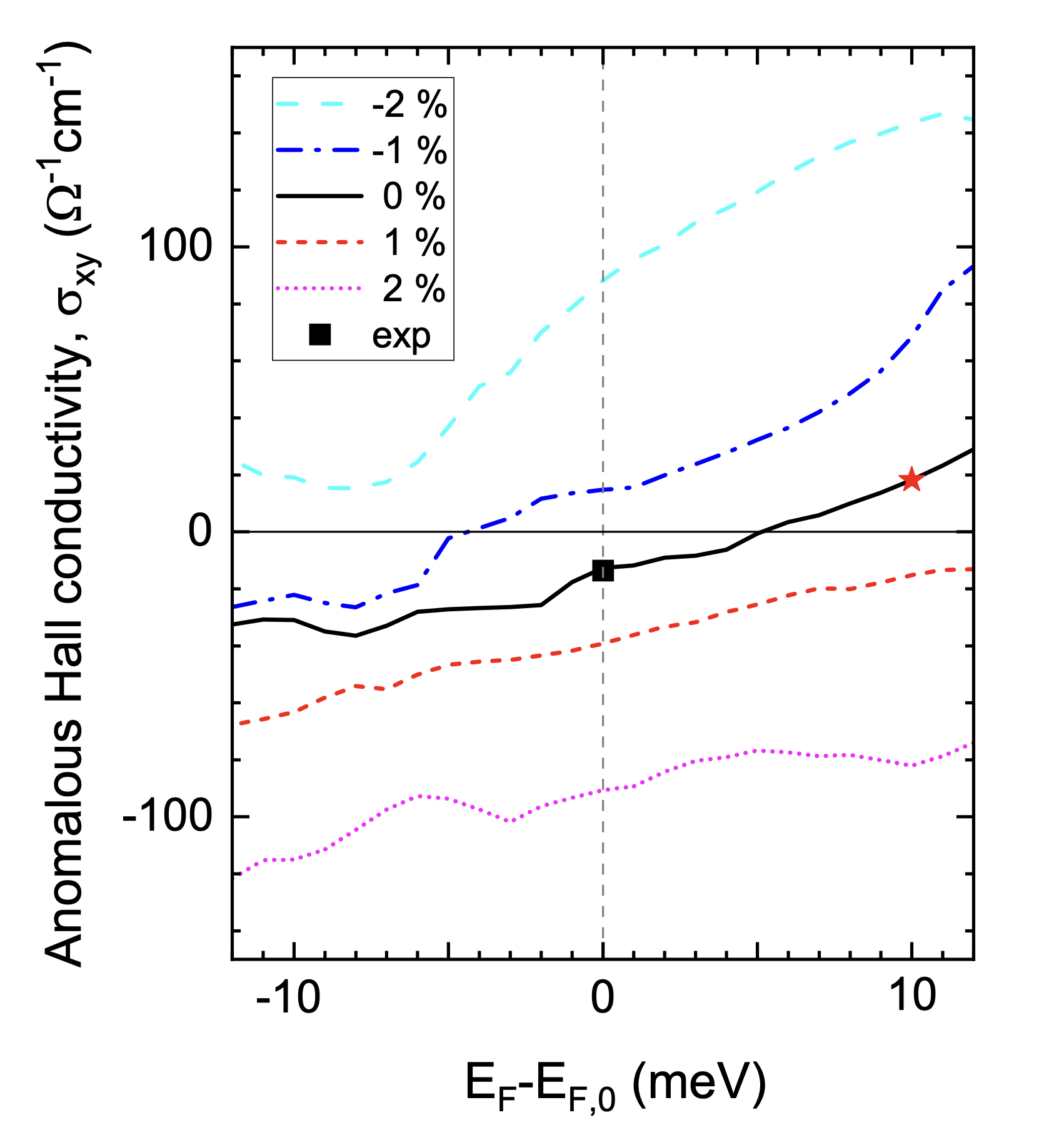}
\caption{Anomalous Hall conductivity as a function of the Fermi energy for different 
bi-axial strains, as indicated in the legend.
Negative (positive) values are for compressive (tensile) strain.
The black square is the experimental value for the AHC in a bulk-like \CT sample 
(24 u.c.) from Ref. \cite{2023_our_NatComm}.
The point at $E_F - E_{F,0} = 10$ meV indicated by the red star is discussed in the text.
}
\label{fig_AHC}
\end{figure}

%
The calculated values of the AHC as a function of the Fermi level with respect to the equilibrium Fermi level
for different values of biaxial strain are shown in Fig. \ref{fig_AHC}.
There exists a general trend that the AHC becomes more positive with filling
for all strains.
For the unstrained case, we will analyze the effect of filling
by analyzing the band structure and Berry curvature at the two
points indicated by the black square and red star in Fig. \ref{fig_AHC}.
For 0\% strain, the value of the intrinsic AHC is $-12.7 \; \Omega^{-1}{\rm cm}^{-1}$, which is in close agreement
with the experimentally measured value of $-11.5 \; \Omega^{-1}{\rm cm}^{-1}$ \cite{2023_our_NatComm}.
As the strain changes from +1\% to $-$1\%, the sign of the 
AHC switches from negative to positive.
These results are in agreement with recently reported experiments 
\cite{2023_our_NatComm,Fujisawa2022,2023_arxiv_Cr2Te3_on_2D_exp_DFT}.

A common metric used to quantify the magnitude of the AHC is the anomalous Hall angle (AHA).
The strain dependent anomalous Hall angles for different strains are shown in Table \ref{table_AHA}.
The AHA is determined by the relative contribution of the anomalous Hall current with respect to the normal current, expressed by
$\Theta_{AH}=\frac{\sigma^{AH}_{xy}}{\sigma_{xx}}$.
We obtained the experimental $\rho_{xx}$ values from \cite{2023_our_NatComm} for a 3 
unit cell (u.c.) thick film under compressive strain (0.493 m$\Omega$cm) 
and a 24 u.c. thick film that was unstrained (0.296 m$\Omega$cm).
To calculate the AHA, we use the average of these two values (0.394 m$\Omega$cm) 
for all values of strain to 
eliminate the effect of different film thicknesses on $\rho_{xx}$. 
\begin{table}[htbp]
	\renewcommand\arraystretch{1.25}
	\caption{\label{table_AHA} Calculated anomalous Hall angle. Experimental longitudinal resistivity $\rho_{xx}$ is obtained from \cite{2023_our_NatComm} }
	\begin{ruledtabular}
		\begin{tabular}{cccc}
			Strain & -1 $\%$ & 0 $\%$ & 1 $\%$\\
			\hline
			$|\sigma^{AH}_{xy}|$ ($\Omega^{-1}cm^{-1}$)	 & 14.7 & 12.7 & 39.2 \\
			\hline
			$\Theta_{AH}$ ($\%$)	 & 0.58 & 0.50 & 1.55 \\
		\end{tabular}
	\end{ruledtabular}
 \label{tab:AHA}
\end{table}
The calculated values of the AHA are comparable to that of  bulk 
Cr$_5$Te$_8$ \cite{2022_Cr5Te8_colossalAHE_AHA_natureElectronics} 
but relatively low among other magnetic materials such as Fe$_3$GeTe$_2$ \cite{2018_LargeAHA_FGT_NatureMater}.

\begin{figure*}[htbp]
\includegraphics[width=0.8\textwidth]{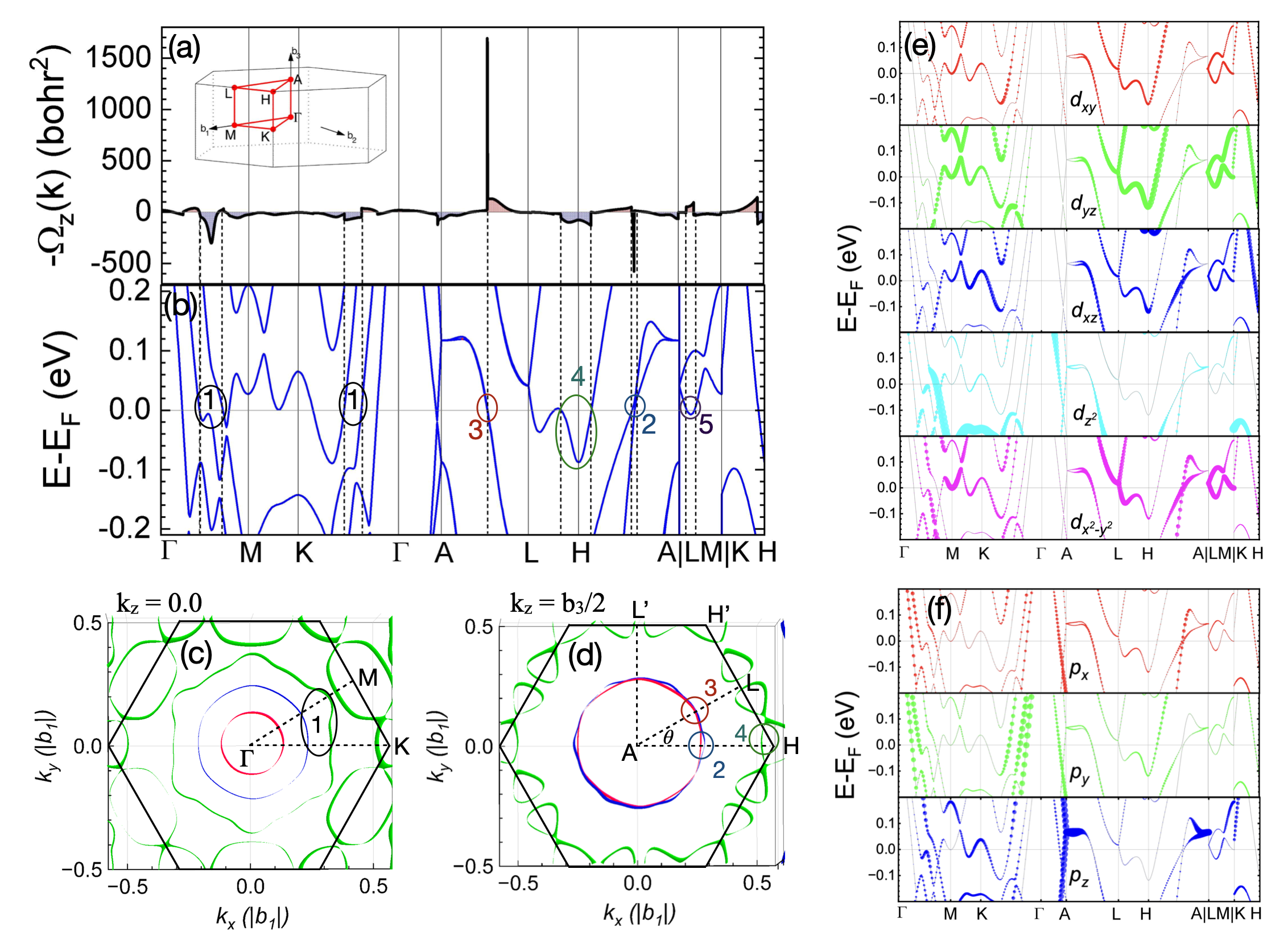}
\caption{
(a) Berry curvature calculated along the high symmetry lines corresponding to the 
(b) electronic dispersion.
Cross sections of the Fermi surface are shown in (c) for $k_z = 0$ (through $\Gamma$)
and (d) for $k_z = b_3/2$ (the top of the Brillouin zone).
The three bands are color coded.
The numbered circled regions correspond to the same numbered
circled regions in (b). 
(e) Cr d-orbital projections of the bands.
(f) Te p-orbital projections of the bands.
}
\label{fig:Ef_Berry_overview}
\end{figure*}
\begin{figure*}[htbp]
\includegraphics[width=0.95\textwidth]{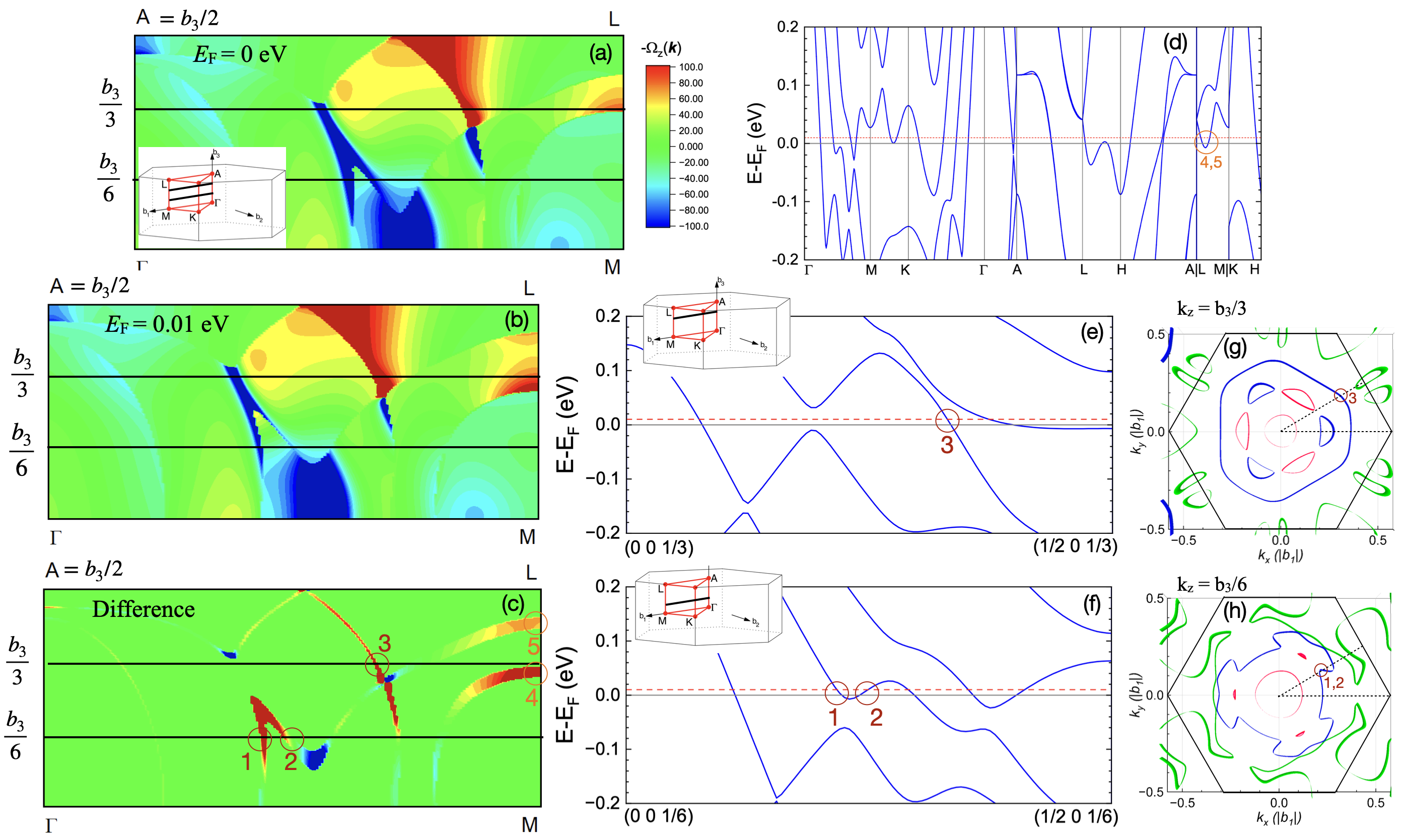}
\caption{
(a,b) Berry curvature contour plots in the two-dimensional $\Gamma$-A-L-M $k$-plane 
for the equilibrium state ($E_F = 0$) (a) and $E_F = 10$ meV (b).
(c) Difference of the Berry curvature plots (a) and (b), 
$\Omega_z(k)(E_F$(10meV)-$E_F$(0meV)).
(c) Electronic band structure along the high symmetry $k$-paths for 0 \% strain. 
The red dashed line shows the Fermi-level shifted upwards by 10 meV.
The orange circle labeled 4,5 corresponds to the circles labeled 4 and 5 in (c).
(e) Band structure calculated along the upper solid lines in (a-c) at 
$k_z = b_3/3$. The path in $k$-space is shown in the inset.
(f) Band structure calculated along the lower lines in (a-c) at
$k_z = b_3/6$. The path in $k$-space is shown in the inset.
The $x$-axis labels in (e) and (f) are in units of the reciprocal 
lattice vectors.
(g,h) Two dimensional slices of the Fermi surface at a fixed $k_z$
as indicated by the labels on each plot. 
The labeled red circles in (g) and (h) correspond to the circles
with the same labels in (e) and (f).
}
\label{fig_shift}
\end{figure*}
%
%
\begin{figure*}[htbp]
\includegraphics[width=0.8\textwidth]{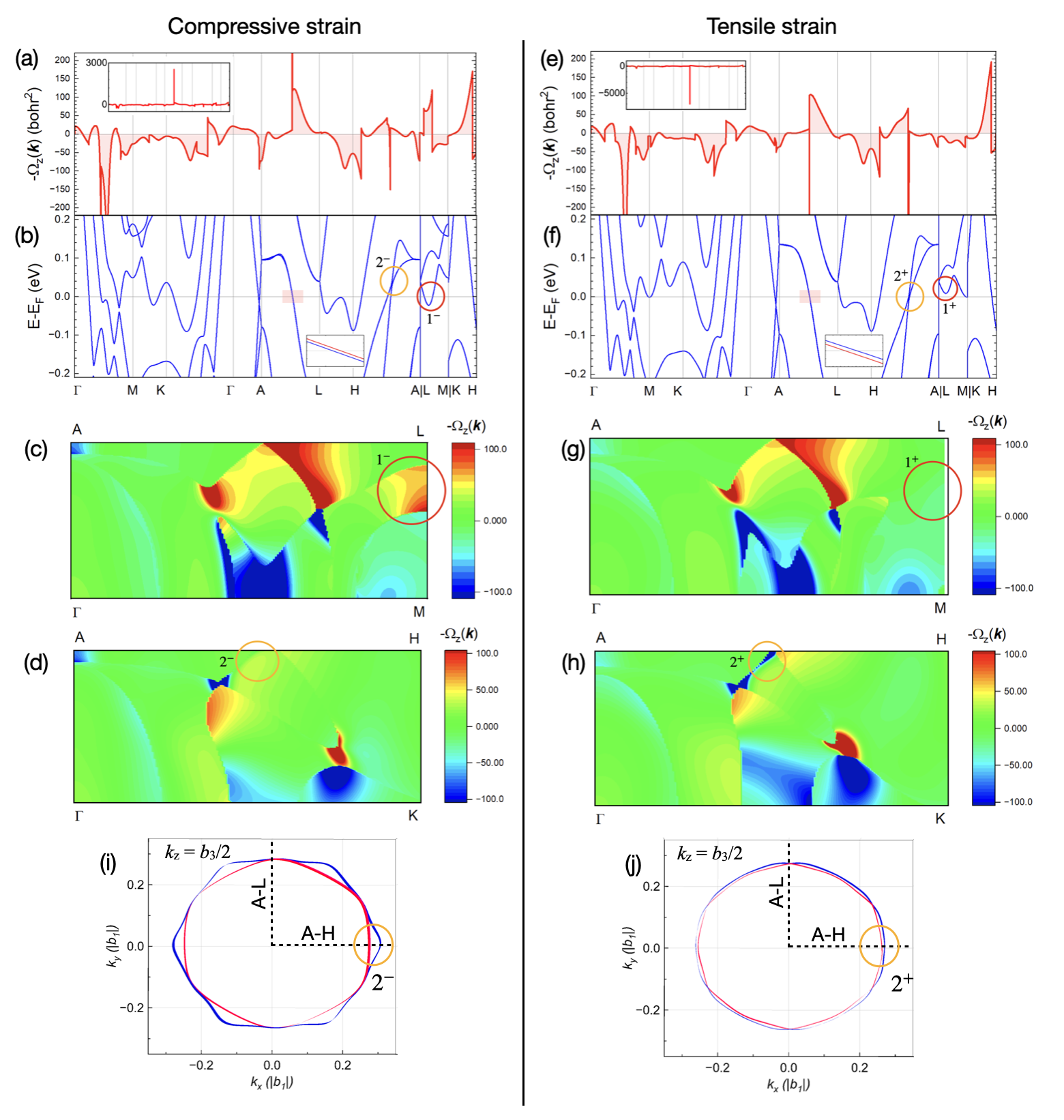}
\caption{
All of the results in the left panel are calculated under 1\% 
compressive strain, and all of the results in the right panel are calculated under
1\% tensile strain.
(a,e) Berry curvature along high symmetry lines of the Brillouin zone.
The vertical axes are truncated for clarity.
The top insets in (a,e) show the same plots over the 
full range of the Berry curvature.
(b,f) Electronic band structure plotted along the same high symmetry lines 
as the Berry curvature. 
The bottom insets in (b) and (f) show the zoomed-in area of highlighted 
rectangular regions in the middle of the A-L $k$-path.
Berry curvature plotted in the two-dimensional planes defined by the 
high symmetry lines (c,g) $\Gamma$-A-L-M and (d,h) $\Gamma$-K-H-A. 
The two planes are illustrated in the inset of 
Fig. \ref{fig_shift}(f).
The color bars show the signs and magnitudes of the Berry curvatures.
(i,j) Zoom in of the two inner rings in the cross sections of the Fermi surfaces under compressive
and tensile strain taken at the top of the Brillouin zone ($k_z = b_3/2$).  
}
\label{fig_Berrys}
\end{figure*}

Before analyzing how the Berry curvature changes with filling and strain, it is useful to first 
understand the relationship between the Berry curvature and the band structure of \CT
and determine which bands make significant contributions.
From Eq. (\ref{eq:Omega-n-z-1}), 
it is clear that $\Omega_n^z(\kv)$ is antisymmetric in $n$ and $n'$.
Therefore, only pairs of occupied and unoccupied bands contribute to the 
sum over $n$ and $n'$ leading to $\Omega^z$.
Furthermore, due to the denominator in (\ref{eq:Omega-n-z-2}), 
two bands closely spaced in energy on either side of the Fermi energy
result in a strong peak in the Berry curvature and a large contribution to $\Omega^z$.
The sensitivity of the Berry curvature to small changes in the
Fermi level further directs our attention to the bands near the Fermi level.

The electronic band structure along the high-symmetry lines and the Berry curvature calculated along the
same lines are shown in Fig. \ref{fig:Ef_Berry_overview}.
One thousand $k$-points are used for each line segment between high symmetry points
to resolve the sharp features in the Berry curvature.
The shaded regions delineate the boundaries where the Berry curvature changes sign. 
These boundaries occur where a band crosses the Fermi level such that an
unoccupied band becomes occupied or vice-versa. 
Since the integral over the entire Brillouin zone of the Berry curvature gives the AHC, which is rather small,
the positive and negative regions of Berry curvature largely cancel.
Thus, relatively small changes in any one region of the Brillouin zone can potentially 
cause a sign change in the AHC. 

Two-dimensional slices of the Fermi surface at $k_z = 0$ (through $\Gamma$)
and $k_z = b_3/2$ (the top of the Brillouin zone) are shown in 
Fig. \ref{fig:Ef_Berry_overview}(c,d).
Cross sections of the Fermi surface over the top half of the Brillouin zone
in steps of $\Delta k_z = b_3/20$ are shown in the Appendix \ref{sec:Fermi_slices} 
in Fig. \ref{fig:Fermi_slices}.
The plots are created by calculating the electronic band structure
on a $400\times 400$ $k_x - k_y$ grid and plotting $E_n(\kv)$
limited to the range $E_F \pm 2$ meV. 
Note that the line widths of these curves 
are artifacts of the plotting software and have no 
physical meaning.
The Fermi surface throughout the Brillouin zone consists of three bands.
Of the 152 bands in the Wannier basis, only bands 115-117 cross the Fermi energy 
somewhere in the Brillouin zone.
These 3 color coded bands comprise the Fermi surface.
Each color, red, blue, and green, identifies a single band
and the corresponding band energies we designate as $E_r(\kv)$, $E_b(\kv)$, and $E_g(\kv)$,
respectively.
The bands are ordered such that at each $\kv$, $E_r < E_b < E_g$.
For example, in the $\kv$ region of the blue band, 
the red band lies below $E_F$ and the green band is above $E_F$.
When a band has multiple non-connected parts, it is a result of the band
having multiple valleys.
At the top of the Brillouin zone (Fig. \ref{fig:Ef_Berry_overview}(d)), 
there are two circular inner bands split by SOC. 
In the orbital projection plots (Fig. \ref{fig:Ef_Berry_overview}(e,f)),
these are the bands crossing the Fermi level along the A-L and A-H lines.
They are primarily Cr d-orbital bands, although they do have some Te p-orbital component.
At $k_z = 0$, the orbital weight of these two bands becomes primarily $p_x$, $p_y$ 
from the Te p-orbital bands.
In the orbital projection plots, these are the first two bands crossing $E_F$
along the $\Gamma$-M and $\Gamma$-K lines.
The third band near the edge of the Brillouin zone in Fig. \ref{fig:Ef_Berry_overview}(d) 
retains a large d-orbital component at $k_z = 0$, 
however, the component changes from
$d_{yz}$ at the top of the Brillouin zone to primarily $d_{z^2}$, $p_z$ at $k_z = 0$.

The two SOC split inner bands in Fig. \ref{fig:Ef_Berry_overview}(d)
give rise to an intense ring of Berry curvature.
The splitting of the two bands
depends on the angle with respect to the $k_x$ axis with a $60^\circ$ rotational periodicity. 
The splitting is largest along the A-H line, since the Fermi level lies just
below the anti-crossing within the number 2 circle in Fig. \ref{fig:Ef_Berry_overview}(b). 
The splitting is smallest along the A-L line where the two bands are split in energy by
3 meV where they cross the Fermi level within the number 3 circle in  Fig. \ref{fig:Ef_Berry_overview}(b).
These closely split bands give rise to the large positive and negative narrow peaks in the 
Berry curvature along the A-L and A-H lines, 
respectively, shown in Fig. \ref{fig:Ef_Berry_overview}(a).
In the two dimensional Brillouin zone, these peaks form a ring of intense Berry curvature 
that alternates in sign every $30^\circ$.
Another region of broad negative Berry curvature arises from the filled local minimum of the 
third band enclosed in the number 4 circle in Fig. \ref{fig:Ef_Berry_overview}(b,d).
As $k_z$ decreases from the top of the Brillouin zone to $\Gamma$, the two inner bands diverge at the
Fermi level in $k$-space,
and the third band has multiple valleys with multiple crossings of the Fermi level.
There is one region of negative Berry curvature that lies in the region of the filled second band and empty
third band, shown in the number 1 circles in Fig. \ref{fig:Ef_Berry_overview}(b,c).
One other region that contributes to both the filling and strain dependence is the small valley
between L and M enclosed in circle $5$ in Fig. \ref{fig:Ef_Berry_overview}(b).
If it depopulates, as it does under tensile strain, its contribution disappears.
With this understanding of the basic features of the band structure and Berry curvature, we 
now discuss the changes with respect to filling and strain.

\subsection{Evolution of the Berry curvature with filling}
\label{sec:filling_Berry_Curv}
 
First, we analyze the effect of band filling on the Berry curvature.
We focus on the Berry curvature in the $\Gamma$-M-L-A plane
with 0\% strain at $E_F = 0$ and $E_F = 10$ meV, 
corresponding to the black square and red star in Fig. \ref{fig_AHC}, 
respectively.
The Berry curvature plots are shown in Fig. \ref{fig_shift}(a) and (b), respectively.
Since the geometric structures and Brillouin zones are identical in both cases,
we can take the difference of the Berry curvature plots for the two different
Fermi energies and plot it,
as shown in Fig. \ref{fig_shift}(c).
To analyze the `hot' regions in the difference plot, we calculate the 
electronic band structure along the paths in the Brillouin zone
indicated by the black solid lines in Figs. \ref{fig_shift}(a-c).
The band structure for the upper line is shown in Fig. \ref{fig_shift}(e),
and the band structure along the lower line is shown in Fig. \ref{fig_shift}(f).
The black line in the inset of both figures shows the $k$-space path 
within the 3D Brillouin zone.
Slices of the Fermi surface at these same two $k_z$ values are shown
in Fig. \ref{fig_shift}(g,h).

We now analyze the regions of large difference in Fig. \ref{fig_shift}(c) using
the band structures calculated along the lines shown.
First, consider the lower line in Fig. \ref{fig_shift}(c).
It intersects the red difference regions at the circles labeled 1 and 2.
These same regions are shown in the band structure in Fig. \ref{fig_shift}(f)
and the Fermi surface in Fig. \ref{fig_shift}(h).
The increase in the Berry curvature arises from a filling of the 
local electron-like minimum at the Fermi level.
The filling of the states on either side of the minimum, lying between the 
red-dashed and solid black lines,
allows them to contribute 
to the sum in Eq. (\ref{eq:Omega-n-z-2})
as a filled state with a matrix element to an empty state that is
closely spaced in energy.
Next, consider the upper line in \ref{fig_shift}(c). 
It intersects the red difference region at the circle labeled 3.
This same region is shown in the band structure in Fig. \ref{fig_shift}(e)
and the Fermi surface in Fig. \ref{fig_shift}(g).
The increase in the Berry curvature arises from the filling of the 
lower band near an anti-crossing, so that there is an unoccupied band
very close in energy.
The final two circled regions 4 and 5 lie along the M-L path shown in 
Fig. \ref{fig_shift}(d).
These two regions also arise from the filling of a local minimum
that is close in energy to an unoccupied band, so that the additional
filled states on either side of the minimum give rise to a large
contribution to the Berry curvature.
We will see that this is the same band that becomes unoccupied under tensile 
strain shown in Fig. \ref{fig_Berrys}(f).

\subsection{Evolution of the Berry curvature with strain}
\label{sec:strain_Berry_Curv}

Now, we consider the effect of strain
on the Berry curvature and the electronic band structure.
Figure \ref{fig_Berrys} shows the Berry curvature and electronic band structures of bulk Cr$_2$Te$_3$ 
along the high symmetry $k$-paths for compressive ($-$1\%) and tensile (+1\%) strains.
The vertical axes of the main plots in (a,e) have been truncated
so that other features can be observed, but the full plots are
shown in the upper insets.
As shown in Fig. \ref{fig_Berrys} (a,e),
the sharp, narrow peak in the Berry curvature 
midway along the A-L line
switches sign under tensile strain.
The sign of the peak along the A-H line remains unchanged.
Thus, the ring of intense Berry curvature along the top of the Brillouin
zone no longer alternates in sign under tensile strain, but it is entirely negative.
This sign change is consistent with the sign change of the AHC with strain.
The bottom insets of Fig. \ref{fig_Berrys}(b,f) show a zoomed-in
region near the Fermi level
that illustrates the underlying mechanism for the
change of sign.
The change in strain causes the order of the two bands to switch, 
as denoted by the red and blue colors.
We determine this by considering the inner product of the eigenvectors
of $H^W(\kv)$ 
corresponding to the two bands 
near the Fermi energy under compressive 
and tensile strain.
There are 4 eigenvectors ($\ket{1,n}$, $\ket{2,n}$, $\ket{1,p}$, $\ket{2,p}$)
where the numbers 1 and 2 indicate the lower and upper band, respectively,
and the letters $n$ and $p$ indicate compressive and tensile strain, respectively.
We find that $|\braket{1n}{2p}|^2 = 0.57$ and $|\braket{1n}{1p}|^2 = 0.38$.
Thus, the lower energy band under compressive strain has a greater overlap
with the upper energy band under tensile strain.
This indicates a switching of the bands under strain.
Since the numerator in the Berry curvature expression in Eq. (\ref{eq:Omega-n-z-1}) 
is antisymmetric in the band index, a reversal of the bands results in a change of sign of the resulting Berry curvature.
This effect can also be seen in the cross-sections of the Fermi surfaces 
under compressive and tensile strain shown in Fig. \ref{fig_Berrys} (i) and (j), respectively.
Under compressive strain, the two bands in (i) never cross, whereas
under tensile strain, the bands in (j) cross at each A-L line.

The peak in the Berry curvature along the A-H line is caused by the 
band crossing in the circles labeled $2^{\mp}$ in Fig. \ref{fig_Berrys}(b,f) and (i,j).
As the strain changes from compressive to tensile, this band crossing moves
closer to the Fermi level, which increases the magnitude of the local Berry curvature.
The intrinsic AHC results from the integration of the Berry curvature 
over the entire Brillouin zone, and, while these peaks have large magnitude, 
they are also very narrow.
Therefore, we now consider the Berry curvature in two different 
two-dimensional slices of the Brillouin zone.

The Berry curvature in the $k$-space planes bounded by
the high symmetry lines 
$\Gamma$-M-L-A and $\Gamma$-K-H-A are shown in Fig. \ref{fig_Berrys}(c,d,g,h).
%
Strain changes the size of the unit cell and thus the Brillouin zones.
Therefore, we are unable to take the difference of the two plots, as we did for filling,
and instead make several observations based on visual comparison.
Comparing the plots for compressive strain on the left to 
the ones for tensile strain on the right, there are two notable changes in the 
Berry curvature.
There is a large positive Berry curvature 
along the M-L line under compressive strain that disappears under tensile strain,
as shown in the red circled areas in Fig. \ref{fig_Berrys}(c,g).
This is a result of the occupied band along M-L under compressive strain, 
shown in the $1^-$ circle in Fig. \ref{fig_Berrys}(b), 
becoming unoccupied under tensile strain, 
shown in the $1^+$ circle in \ref{fig_Berrys}(f).
Another contribution to the change in sign of the AHC comes from the band crossing,
discussed above,
circled in the $2^{\mp}$ circles in Fig. \ref{fig_Berrys}(b,f). 
Under compressive strain, the band crossing occurs above the Fermi level,
and under tensile strain, the crossing occurs closer to the Fermi level. 
This results in a large negative contribution to the Berry curvature near the A-H 
line shown in the region enclosed by $2^+$ circle in 
Fig. \ref{fig_Berrys}(d,h).
Thus, the change of sign of the AHC with strain is consistent with the shifts
of the bands near the Fermi level, which either reduce the positive Berry curvature
or create the regions of negative Berry curvature.

\section{Conclusion}

In bulk \CT, 
we conclude that the sensitivity of the sign
of the Berry curvature to strain and filling arises
from the three bands that cross the Fermi level.
There are multiple regions of positive and negative Berry curvature 
throughout the Brillouin zone that mostly cancel, so that small changes
in the local Berry curvature can cause an overall sign change in the total integrated value,
which determines the AHC.
The intense ring of Berry curvature at the top of the Brillouin zone
associated with the two SOC split bands is most striking.
In the absence of tensile strain,
the sign of the Berry curvature associated with the ring oscillates with each
$30^\circ$ rotation around the ring.
Tensile strain causes a reversal of the band order along the A-L line such that
the Berry curvature of the ring becomes uniformly negative.
Other regions also contribute, and one that we have particularly identified
comes from the local minimum along the M-L line at the face of the Brillouin zone.
Tensile strain causes it to depopulate, which removes its positive contribution to the 
Berry curvature.
The analysis of band filling shows that the changes in the Berry curvature, which
drive the sign change in the AHC, occur
throughout the Brillouin zone
and that the largest changes in Berry curvature do not come from the ring.
Filling (or depopulating) of local minima close to the unoccupied bands, or filling of 
bands near anti-crossings, 
gives rise to the changes in the local Berry curvature
that drive the sign change of the AHC with filling.
In all cases, the contributions come from the change in filling of bands
in close energetic proximity to unoccupied bands.
In thin films, termination with the Te layer gives the most stable structures,  
although thin films terminated with Cr are also stable.
Terminating the surface at the partially occupied Cr layer increases the in-plane lattice
constant, and it increases the spin polarization of the electrons at the Fermi level.
Only when both surfaces are terminated at the partially occupied Cr layer, 
do the thin films become half metals in 1 layer and 2 layer films.
The Cr termination serves as a donor dopant moving the Fermi level into the minority spin
gap, and it also increases the magnitude of the minority spin gap.
%

\begin{acknowledgments}
This work was supported in part by the U.S. Army Research Laboratory (ARL) Research Associateship Program (RAP) 
Cooperative Agreement(CA) W911NF-16-2-0008.
This work used STAMPEDE2 at TACC through allocation DMR130081 from the Advanced Cyberinfrastructure Coordination Ecosystem: 
Services $\&$ Support (ACCESS) program, 
which is supported by National Science Foundation grants $\#$2138259, $\#$2138286, $\#$2138307, $\#$2137603, and $\#$2138296.
\end{acknowledgments}

\appendix
\section{Phonon Dispersion and Molecular Dynamics Calculations}
\label{app:phonon_disp_MD}
\begin{figure*}[htpb]
	\centering
	\includegraphics[width = 0.9\textwidth]{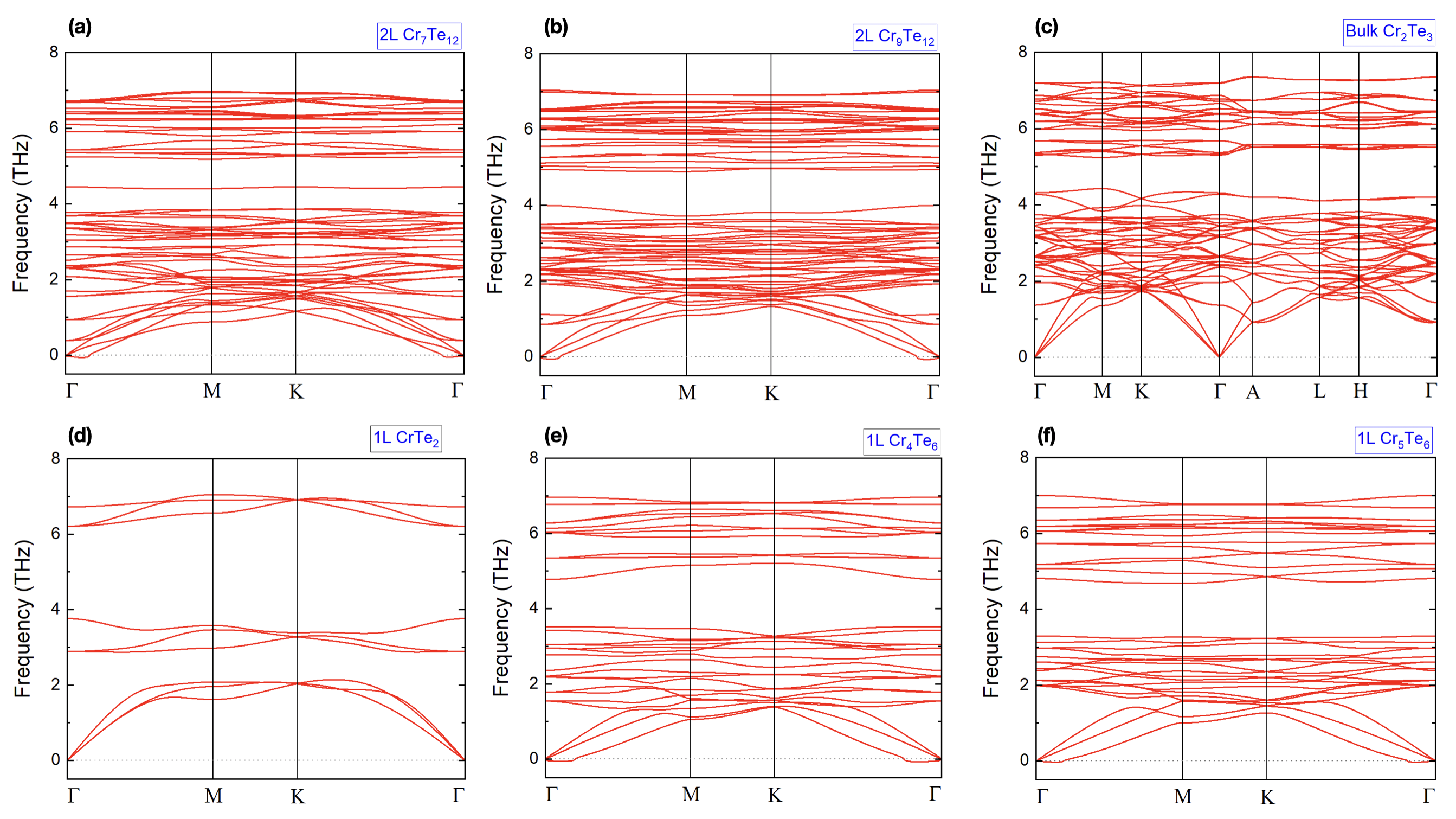}
	\caption{Phonon dispersions of (a) 2L Cr$_7$Te$_{12}$ (b) 2L Cr$_9$Te$_{12}$ (c) bulk Cr$_2$Te$_3$ (d) 1L CrTe$_2$ (e) 1L Cr$_4$Te$_6$ and (f) 1L Cr$_5$Te$_6$.}
	\label{fig:phonon}
\end{figure*}

In order to confirm the structural stability of bulk and thin films of \CT, we calculate the 
phonon dispersion.
A 2$\times$2$\times$2 supercell with a $k$-grid of 4$\times$4$\times$2 is used for the bulk and a 2$\times$2$\times$1 supercell with a $k$-grid of 6$\times$6$\times$1 is chosen for thin films.
All thin film unit cells contain a 15-\AA-thick vacuum region.
Figure \ref{fig:phonon}(c) shows the phonon dispersion for bulk \CT, Fig. \ref{fig:phonon}(a,b) shows
the phonon dispersions for bilayer thin films, and Fig. \ref{fig:phonon}(d-f) shows the phonon dispersions for monolayer thin films.
The phonon dispersion of (d) CrTe$_2$ was calculated by us previously, and it requires
a Hubbard parameter of $U = 2$ eV to obtain a dispersion with non-negative 
frequencies (see the Supplemental Material of \cite{Yuhang_CrTe2}).
All of the other plots are calculated with $U=0$.
The phonon dispersions of all of these other few-layer structures show small regions of
negative frequency near $\Gamma$, which is not unusual for these types of calculations.

\begin{figure*}
    \includegraphics[width = 0.85\textwidth]{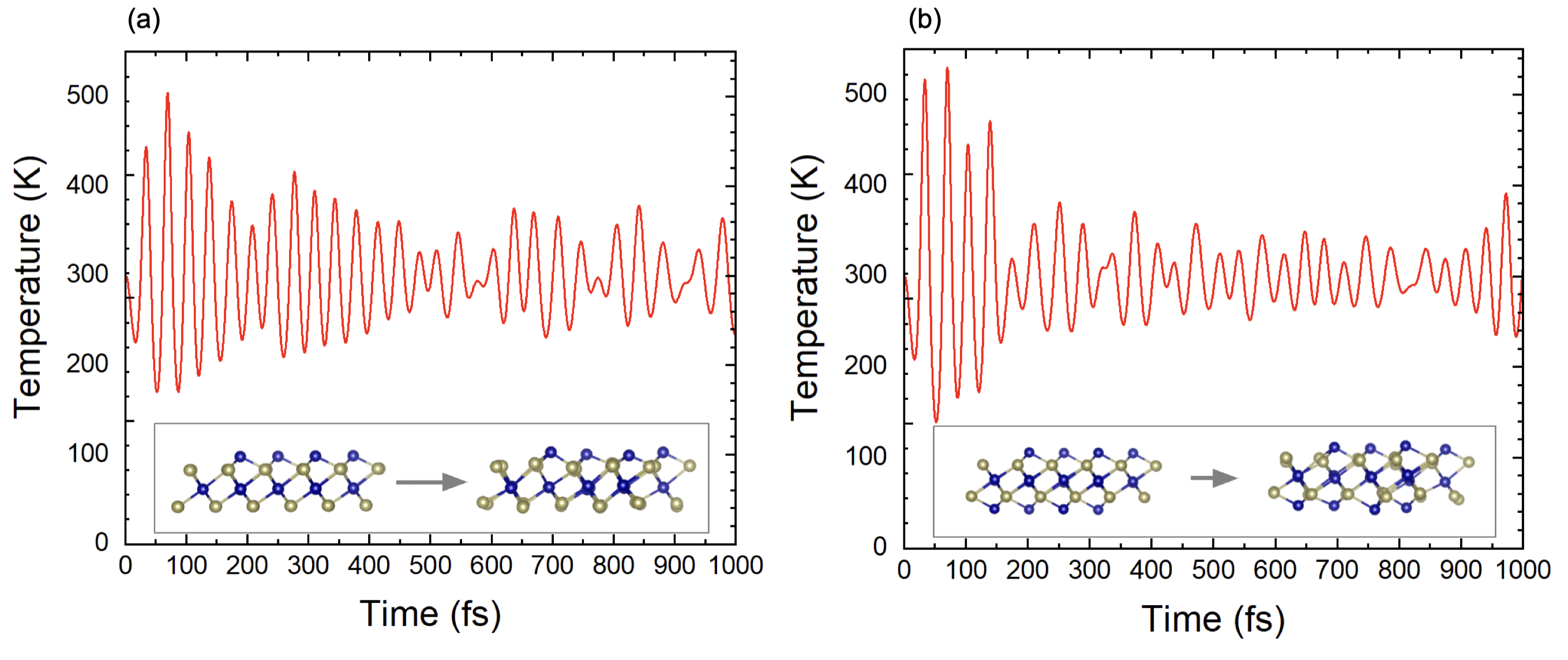}
    \caption{Constant temperature AIMD calculations for Cr/Te-terminated Cr$_4$Te$_6$ (a) and Cr/Cr-terminated Cr$_5$Te$_6$ (b). 
    Red curve represents the temperature as a function of time. Insets show the crystal structures at 0 fs and 1 ps for each calculation.}
\label{fig:MD}
\end{figure*}

Therefore, to further verify the stability of the thin films, especially with Cr termination, 
we perform AIMD
calculations, as implemented in VASP.
Using 2$\times$2$\times$1 supercell and a temperature of 300 K, we perform AIMD calculations
with a 1 fs time step over a period of 1 ps.
Figure \ref{fig:MD} shows temperature oscillations as a function of time 
at a constant average temperature of 300 K for Cr/Te- and Cr/Cr- terminated monolayer thin films.
%
%
The insets of Fig. \ref{fig:MD} show crystal structures at the beginning and the end of calculations.
The final structures are slightly distorted, however, they maintain the same crystal structure, 
and all of the bonds are intact. 
These results indicate that both Cr$_4$Te$_6$ and Cr$_5$Te$_6$ thin films are thermodynamically stable.

\section{Method Details}
\label{k-convergence}
\ref{table:kmesh} shows the results of the $k$-grid convergence test for the 
calculation of the AHC.
Table \ref{tab:U} compares the experimentally known values of the
magnetic moment and AHC with values calculated
with $U=0$ and $U=2$ eV.

\begin{table}[htbp]\centering
\caption{\label{table:kmesh} 
 Convergence of the AHC as a function of the $k$-grid.}
		\begin{tabular}{ccc}
			$k$-mesh & Adaptive refinement & AHC ($\Omega^{-1}cm^{-1}$) \\
			\hline \hline
			40x40x40 & - & -9.53 \\
			\hline
			40x40x40 & 3x3x3 & -19.24 \\
            \hline
			50x50x50 & 3x3x3 & -12.65 \\
            \hline
			80x80x80 & - & -12.67 \\
            \hline
			100x100x100 & - & -12.7 \\
		\end{tabular}
\end{table}

\begin{table}[htbp]\centering
\caption{Experimental values \cite{2023_our_NatComm} and calculated average magnetic moments and anomalous Hall conductivity for $U=0$ and $U=2$ eV.} 
		\begin{tabular}{c|c|cc}
            & Experimental & $U=0$ eV & $U=2$ eV \\
			\hline \hline
			Magnetic moment ($\mu_B$) & 2.84 & 2.85 & 3.1 \\
			\hline
			AHC ($\Omega^{-1}cm^{-1}$) & -11.5 & -12.7 & -120.95 \\
           
		\end{tabular}
 \label{tab:U}
\end{table}

\section{Slices of the Fermi Surface at Fixed $k_z$}
\label{sec:Fermi_slices}
\begin{figure*}
    \includegraphics[width = 0.7\textwidth]{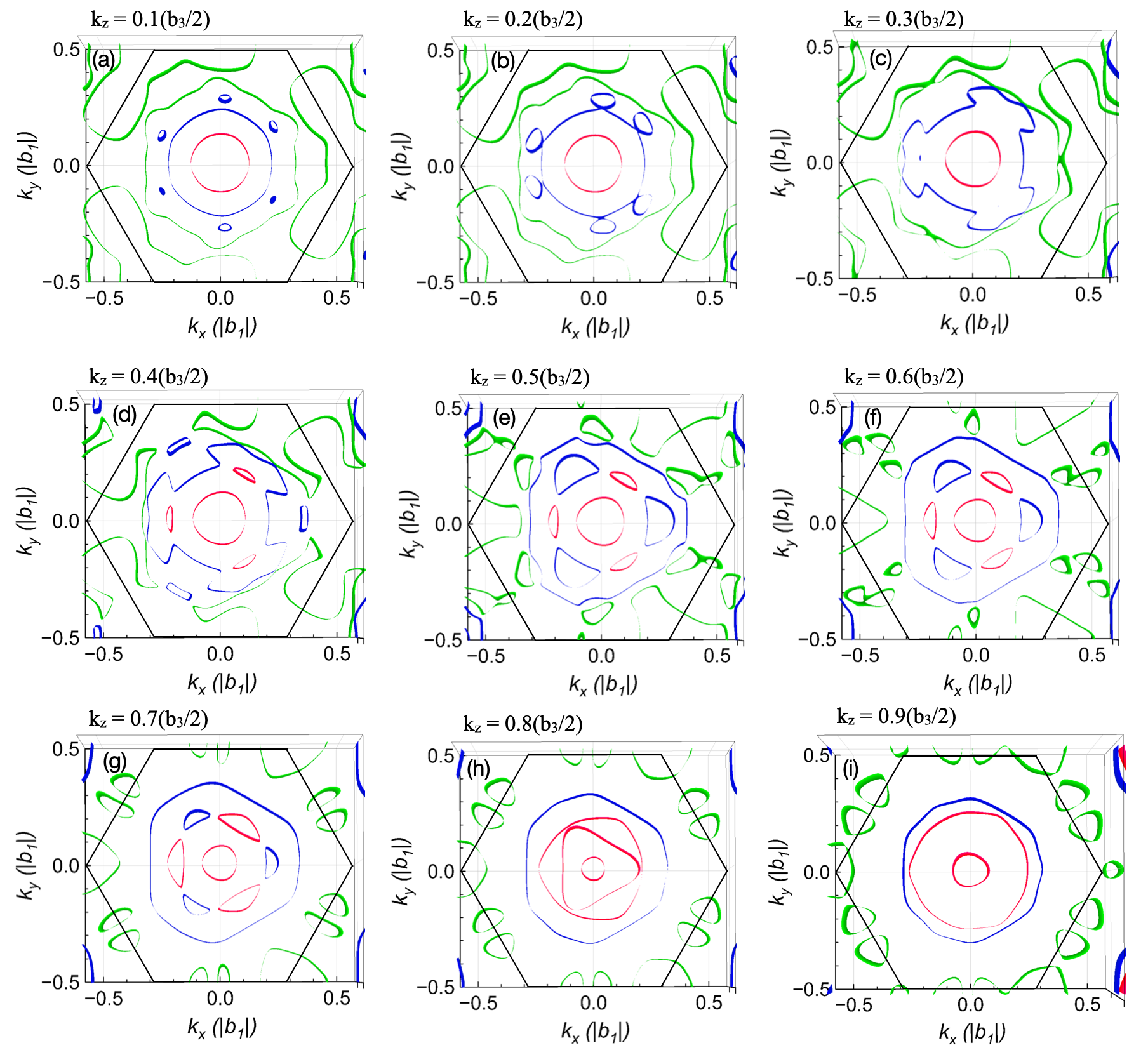}
    \caption{Two-dimensional slices of the Fermi surface at different 
    values of $k_z$ as shown.}
\label{fig:Fermi_slices}
\end{figure*}

Figure \ref{fig:Fermi_slices} shows slices of the Fermi surface of
the top half of the Brillouin zoned at fixed $k_z$
in steps of $\Delta k_z = 0.1 (b_3/2)$. 
The Brillouin zone projected onto the $k_x - k_y$ plane is outlined. 

\clearpage
\bibliography{main.bib}

\end{document}